\documentclass{jkas}

\def\year{2022} 
\def\month{00} 
\def\volume{00} 
\def\issue{0} 
\def\beginpage{1} 
\def\endpage{8} 
\def\received{---} 
\def\accepted{---} 
\date{Received \received; accepted \accepted}

\usepackage{natbib,bm,graphicx}
\usepackage{orcidlink}

\newcommand{\te}{t_{\rm E}}
\newcommand{\tzero}{t_{0}}
\newcommand{\uo}{u_{0}}

\newcommand{\pie}{\pi_{\rm E}}
\newcommand{\pis}{\pi_{\rm S}}
\newcommand{\pien}{\pi_{\rm E,N}}
\newcommand{\piee}{\pi_{\rm E,E}}

\newcommand{\thetae}{\theta_{\rm E}}
\newcommand{\thetas}{\theta_{\rm \star}}

\newcommand{\murel}{\mu_{\rm rel}}
\newcommand{\dl}{D_{\rm L}}
\newcommand{\ds}{D_{\rm S}}

\newcommand{\delcs}{\Delta \chi^{2}}

\newcommand{\fsone}{F_{\rm S,1}}
\newcommand{\fstwo}{F_{\rm S,2}}
\newcommand{\qflux}{F_{\rm S,2}/F_{\rm S,1}}

\title{OGLE-2019-BLG-0362L$\textbf{\rm b}$: A super-Jovian-mass planet around a low-mass star}

\author[1,2,13]{Sun-Ju Chung\, \orcidlink{0000-0001-6285-4528}}
\author[2,13]{Jennifer C. Yee\, \orcidlink{0000-0001-9481-7123}}
\author[3,14]{Andrej Udalski\, \orcidlink{0000-0001-5207-5619}}
\author[4,5,13]{Andrew Gould}
\author[6, 13]{Michael D. Albrow\, \orcidlink{0000-0003-3316-4012}}
\author[1,7,13]{Youn Kil Jung\, \orcidlink{0000-0002-0314-6000}}
\author[1,13]{Kyu-Ha Hwang\, \orcidlink{0000-0002-9241-4117}}
\author[8,13]{Cheongho Han\, \orcidlink{0000-0002-2641-9964}}
\author[1,13]{Yoon-Hyun Ryu\, \orcidlink{0000-0001-9823-2907}}
\author[2,8,13]{In-Gu Shin\, \orcidlink{0000-0002-4355-9838}}
\author[9,13]{Yossi Shvartzvald\, \orcidlink{0000-0003-1525-5041}}
\author[10,13]{Weicheng Zang\, \orcidlink{0000-0001-6000-3463}}
\author[1,11,13]{Sang-Mok Cha}
\author[1,13]{Dong-Jin Kim}
\author[1,13]{Seung-Lee Kim\, \orcidlink{0000-0003-0562-5643}}
\author[1,13]{Chung-Uk Lee\, \orcidlink{0000-0003-0043-3925}}
\author[1,13]{Dong-Joo Lee}
\author[1,11,13]{Yongseok Lee}
\author[1,7,13]{Byeong-Gon Park\, \orcidlink{0000-0002-6982-7722}}
\author[4,13]{Richard W. Pogge\, \orcidlink{0000-0003-1435-3053}}
\author[3,14]{Radek Poleski}
\author[3,14]{Przemek Mr{\'o}z\, \orcidlink{0000-0001-7016-1692}}
\author[3,14]{Pawe\l{} Pietrukowicz\, \orcidlink{0000-0002-2339-5899}}
\author[3,14]{Jan Skowron\, \orcidlink{0000-0002-2335-1730}}
\author[3,14]{Micha\l{} K. Szyma{\'n}ski\, \orcidlink{0000-0002-0548-8995}}
\author[3,14]{Igor Soszy{\'n}ski\, \orcidlink{0000-0002-7777-0842}}
\author[3,14]{Szymon Koz{\l}owski}
\author[3,14]{Krzysztof A. Rybicki}
\author[3,14]{Patryk  Iwanek\, \orcidlink{0000-0002-6212-7221}}
\author[3,14]{Marcin Wrona}
\author[3,14]{Mariusz Gromadzki}
\author[12,14]{Krzysztof Ulaczyk\, \orcidlink{0000-0001-6364-408X}}

\affil[1]{Korea Astronomy and Space Science Institute, 776 Daedeokdae-ro, Yuseong-Gu, Daejeon 34055, Republic of Korea; \email{sjchung@kasi.re.kr}}
\affil[2]{Center for Astrophysics $\mid$ Harvard \& Smithsonian, 60 Garden Street, Cambridge, MA 02138, USA}
\affil[3]{Astronomical Observatory, University of Warsaw, AI.~Ujazdowskie~4, 00-478~Warszawa, Poland}
\affil[4]{Department of Astronomy, Ohio State University, 140 W. 18th Avenue, Columbus, OH 43210, USA}
\affil[5]{Max-Planck-Institute for Astronomy, K{\"o}nigstuhl 17, D-69117 Heidelberg, Germany}
\affil[6]{Department of Physics and Astronomy, University of Canterbury, Private Bag 4800 Christchurch, New Zealand}
\affil[7]{Korea University of Science and Technology, Korea, (UST), 217 Gajeong-ro, Yuseong-gu, Daejeon 34113, Republic of Korea}
\affil[8]{Department of Physics, Chungbuk National University, Cheongju 361-763, Republic of  Korea}
\affil[9]{Department of Particle Physics and Astrophysics, Weizmann Institute of Science, Rehovot 76100, Israel}
\affil[10]{Department of Astronomy and Tsinghua Centre for Astrophysics, Tsinghua University, Beijing 100084, China}
\affil[11]{School of Space Research, Kyung Hee University, Giheung-gu, Yongin, Gyeonggi-do, 17104, Republic of Korea}
\affil[12]{Department of Physics, University of Warwick, Gibbet Hill Road, Coventry, CV4 7AL, UK}
\affil[13]{The KMTNet Collaboration}
\affil[14]{The OGLE Collaboration}

\def\corrauthor{Sun-Ju Chung}

\def\runningauthor{Chung et al.}

\def\runningtitle{A super-Jupiter-mass planet orbiting a low-mass star}

\def\keywords{gravitational lensing: micro}

\def\abstracttext{
We present the analysis of a planetary microlensing event OGLE-2019-BLG-0362 with a short-duration anomaly $(\sim 0.4\, \rm days)$ near the peak of the light curve, which is caused by the resonant caustic.
The event has a severe degeneracy with $\delcs = 0.9$ between the close and the wide binary lens models both with planet-host mass ratio $q \simeq 0.007$.
We measure the angular Einstein radius but not the microlens parallax, and thus we perform a Bayesian analysis to estimate the physical parameters of the lens.
We find that the OGLE-2019-BLG-0362L system is a super-Jovian-mass planet $M_{\rm p}=3.26^{+0.83}_{-0.58}\, M_{\rm J}$ orbiting an M dwarf $M_{\rm h}=0.42^{+0.34}_{-0.23}\, M_\odot$ at a distance $\dl =5.83^{+1.04}_{-1.55}\, \rm kpc$.
The projected star-planet separation is $a_{\perp} = 2.18^{+0.58}_{-0.72}\, \rm AU$, which indicates that the planet lies beyond the snow line of the host star.
}

\begin{document}
\jkashead 

\section{Introduction}
As of 8th April 2022, 130\footnote{https://exoplanetarchive.ipac.caltech.edu/index.html} confirmed exoplanets have been discovered by microlensing.
Host stars of the microlensing planets have wide mass ranges from brown dwarfs to Sun-like stars, and the majority of them are low-mass M dwarf stars, which comprise most of the stars in our Galaxy.
By contrast, host stars of exoplanets detected by the radial velocity and transit methods, which detected over 95\% of 5009$^1$ exoplanets discovered so far, are mostly Sun-like stars.
Using these two methods, it is difficult to detect planets around low-mass M dwarf stars because the stars are faint.
This is because the radial velocity and transit methods depend on the brightness of stars, but the microlensing depends on their mass, not their brightness.
In addition, the great majority of the low-mass host stars detected by microlensing have giant planets beyond their snow lines where ices condense in the protoplanetary disk \citep{kennedy2008}, while Sun-like host stars detected by the two other methods have mostly planets inside the their snow lines, regardless of the planet mass (see Figure 10 of \citet{zhu2021}).
Hence, microlensing planets play a very important role to constrain not only planet formation theories, such as the core accretion and gravitational instability, which were constructed based on observed exoplanets, but also the distribution of exoplanets around all types of host stars.

A planetary signal induced by microlensing is unpredictable and its duration decreases as the planet mass decreases (e.g., a few hours for a Earth-mass planet and about a day for a Jupiter-mass planet).
Thus, for the detection of the microlensing planetary signal, it is highly advantageous  to have 24 hr continuous high-cadence observations.
In order to conduct the 24 hr observations, Korea Microlensing Telescope Network (KMTNet; \citealt{kim2016}) with $4.0\ \rm deg^{2}$ field of view (FOV) was established at three different southern sites, which are located at the Cerro Tololo Interamerican  Observatory in Chile (KMTC), the South African  Astronomical Observatory (KMTS), and the Siding Spring Observatory in Australia (KMTA), and its experiment was officially initiated in 2016.
KMTNet covers 27 Galactic bulge fields at cadences ranging from $\Gamma = 0.2\, \rm hr^{-1}$ to $\Gamma = 4\, \rm hr^{-1}$, with about 12 deg$^2$ at $\Gamma = 4\, \rm hr^{-1}$, 29 deg$^2$ at $\Gamma = 1\, \rm hr^{-1}$,  44 deg$^2$ at $\Gamma = 0.4\, \rm hr^{-1}$, 12 deg$^2$ at $\Gamma = 0.2\, \rm hr^{-1}$, and total coverage of 97 deg$^2$ \citep{shin2016}. 
Thanks to the 24 hr high-cadence observations of KMTNet, many planetary events that were not detected or noticed by the Optical Gravitational Lensing Experiment (OGLE; \citealt{udalski2003}) or the Microlensing Observations in Astrophysics (MOA; \citealt{sumi2016}) were detected by the KMTNet (e.g., \citealt{shin2016}; \citealt{hwang2018a}; \citealt{kim2020}; \citealt{kim2021a}; \citealt{kim2021b}; \citealt{zang2021a}; \citealt{hwang2022}).
Hence, since the beginning of test observations in 2015, about 50\% of all confirmed microlensing exoplanets have been detected by KMTNet.
Considering that it took about 25 years to detect the other 50\% exoplanets before KMTNet, it is obvious how amazing KMTNet is doing.
In addition, very low planet-star mass ratio ($q$) events with $q < 10^{-4}$, which were rarely detected before KMTNet, are often being detected by KMTNet (\citealt{shvartzvald2017}; \citealt{hwang2018b}; \citealt{udalski2018}; \citealt{han2018}; \citealt{gould2020}; \citealt{herrera-martin2020}; \citealt{ryu2020}; \citealt{han2021}; \citealt{kondo2021}; \citealt{zang2021a}; \citealt{zang2021b}; \citealt{yee2021}; \citealt{han2022a}; \citealt{han2022b}; \citealt{hwang2022}; \citealt{wang2022}).
Such events will be helpful to better constrain the planet frequency as a function of the plant-star mass ratio, which was done by \citet{shvartzvald2016} and \citet{suzuki2016}.

Moreover, since the KMTNet microlensing observations, planetary events caused by lens systems that are composed of giant planets beyond the snow line of low-mass host stars are routinely being detected.
The event OGLE-2019-BLG-0362 is one of such events.
In this paper, we report on the discovery of a giant planet orbiting an M dwarf from the analysis of the microlensing event OGLE-2019-BLG-0362.
The mass and distance of lens systems can be directly measured from the measurement of two parameters of microlens parallax $\pie$ and angular Einstein ring radius $\thetae$.
This is because they are defined as
\begin{equation}
M_{\rm L} = {\thetae\over{\kappa\pie}}; \quad \dl = {{\rm AU}\over{\pie\thetae + \pis}},
\end{equation}
where $\kappa \equiv 4G/(c^2\rm AU) \simeq 8.14\, {\rm mas}/M_\odot$, $\pis={\rm AU}/\ds$ is the parallax of the source star, and $\ds$ is the distance to the source \citep{gould2000}.
For the event, only $\thetae$ was measured, thus the physical parameters of the lens system were estimated from a Bayesian analysis \citep{jung2018}.
\citet{kim2021b} showed that the lens masses inferred from Bayesian analyses are determined almost entirely by the measured $\thetae$ and are relatively insensitive to the relative lens-source proper motion $\murel$ and specific Galactic model prior.
\citet{shan2019} also showed that the true distribution of masses for events with measured masses and \textit{Sptizer} parallaxes is consistent with the inferred masses from Bayesian analyses derived for those events.
Therefore, it is believed that the physical lens parameters estimated from the Bayesian analysis are reliable at least in a statistical sense.

\section{Observations}
The microlensing event OGLE-2019-BLG-0362 occurred at equatorial coordinates $({\rm RA, Dec})=(17:33:51.66,-24:48:37.3)$, corresponding to the Galactic coordinates $(l, b) = (2.11^{\circ}, 4.41^{\circ})$.
The event was first detected by the Early Warning System of the OGLE.
OGLE uses a 1.3 m telescope with $1.4\ \rm deg^{2}$ FOV at Las Campanas Observatory in Chile.
The event OGLE-2019-BLG-0362 is located in the OGLE IV field BLG715, which is observed with a low cadence of $\Gamma \sim 1\, \rm night^{-1}$.
The KMTNet also found this event, and it was designated as KMT-2019-BLG-0075.
KMTNet uses three identical 1.6m telescopes with $4.0\ \rm deg^{2}$ FOV, as mentioned in Section~1.
The event is located in the KMT field BLG16 with a cadence of $\Gamma = 0.4\, \rm hr^{-1}$.

Most KMT data were taken in the $I$ band, and some data were taken in $V$ band in order to determine the color of the source star.
The KMT data for the modeling were reduced by the pySIS pipeline based on the difference imaging method (\citealt{alard1998}; \citealt{albrow2009}).
For the construction of the color-magnitude diagram (CMD) of stars around the source and characterization of the source color,  the KMTC $I$- and $V$-band images were used, and they were reduced by the pyDIA pipeline developed by \citet{albrow2007}.
The OGLE data were also reduced by difference imaging analysis (\citealt{alard1998}; \citealt{wozniak2000}).
We renormalized the errors of the data sets obtained from each photometry pipeline using the method of \citet{yee2012}.

\section{Light Curve Analysis}
The event OGLE-2019-BLG-0362/KMT-2019-BLG-0075 has a short duration anomaly ($\sim 0.4\, \rm days$) near the peak.
The anomaly was covered by only three KMT data points (two KMTC and one KMTS data), while it was not covered by OGLE.
In other words, there are no binary lensing features that can be unambiguously identified, such as a clearly double-peaked caustic-crossing.
In such cases, the short duration anomaly can be produced by either a typical binary lensing with a single source (2L1S) or a single lensing with two sources (1L2S).

\begin{figure}
\centering
\includegraphics[width=90mm]{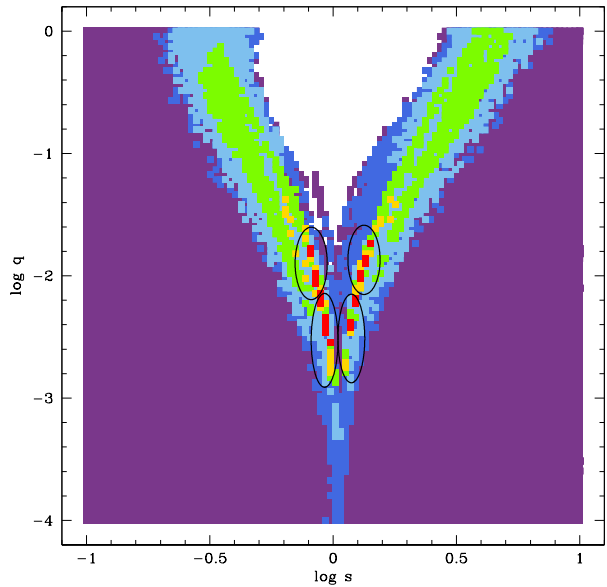}
\caption{Distributions of $\delcs$ in the $(\log s, \log q)$ plane obtained from the grid search.
The colors marked as red, yellow, green, sky blue, vivid blue, and purple represent the regions with $\delcs < 8^{2},\, 16^{2},\, 24^{2},\, 32^{2},\, 40^{2},\, \textrm{and}\,  48^{2}$, respectively.\label{fig:jkasfig1}}
\end{figure}

\subsection{binary lens model (2L1S)}
We first conduct the standard binary lens modeling.
The standard binary lens event is described with seven parameters.
Three of the seven parameters are the single lensing parameters $(\tzero, \uo, \te)$, where $\tzero$ is the time of the closest source approach to the lens, $\uo$ is the impact parameter (i.e., the lens-source separation at $t=\tzero$ in units of $\thetae$), and $\te$ is the Einstein radius crossing time of the event.
Another three parameters are the binary lensing parameters $(s, q, \alpha)$, in which $s$ is the projected separation of the binary lens components in units of $\thetae$, $q$ is the mass ratio of the two lens components, and $\alpha$ is the angle between the source trajectory and the binary axis.
The last one is the normalized source radius $\rho = \thetas/\thetae$, where $\thetas$ is the angular radius of the source star.
Because the short-duration anomaly is likely to be caused by the caustic, we incorporate the limb-darkening variation of the finite source star in the modeling.
The brightness variation of the source by the limb-darkening effect is computed by $S \propto 1 - \Gamma(1-3\cos\phi/2)$, where the $\Gamma$ is the limb-darkening coefficient and $\phi$ is the angle between the normal to the surface of the source star and the line of sight \citep{an2002}.
According to the source type that will be discussed in Section~4 , we adopt the limb-darkening coefficient of $\Gamma_{I} = 0.45$ from \citet{claret2000}.
Besides these parameters, there are two flux parameters for each observatory,  which are the source flux $f_{s,i}$ and blended flux $f_{b,i}$ of the $i$th observatory.
The two flux parameters $(f_{s,i}, f_{b,i})$ are modeled by $F_{i}(t) = f_{s,i}A_{i}(t) + f_{b,i}$, where $A_i$ is the magnification as a function of time at the $i$th observatory \citep{rhie1999}.
Then, the $(f_{s,i}, f_{b,i})$ of each observatory are obtained from a linear fit.

\begin{figure}
\centering
\includegraphics[width=90mm]{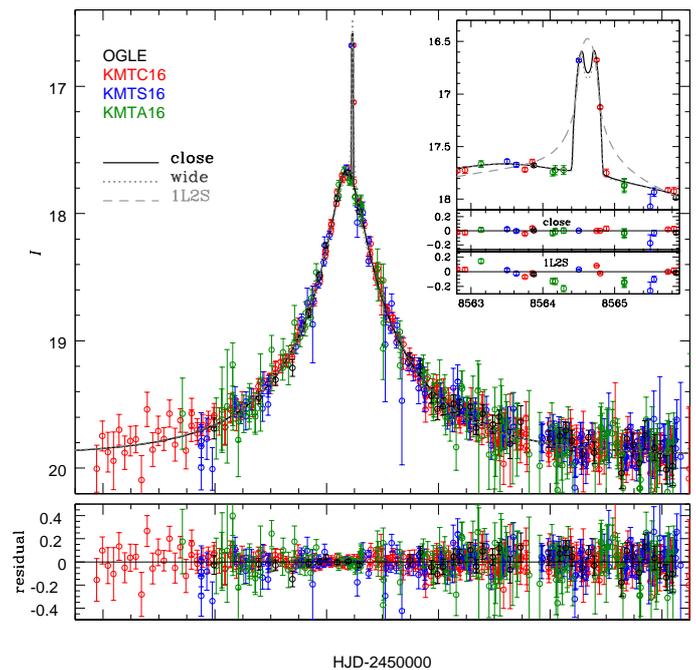}
\caption{Best-fit light curves of OGLE-2019-BLG-0362 for the close and wide binary lens models.
The gray dashed curve is the light curve of the binary source model (1L2S).\label{fig:jkasfig2}}
\end{figure}

\begin{figure}
\centering
\includegraphics[width=90mm]{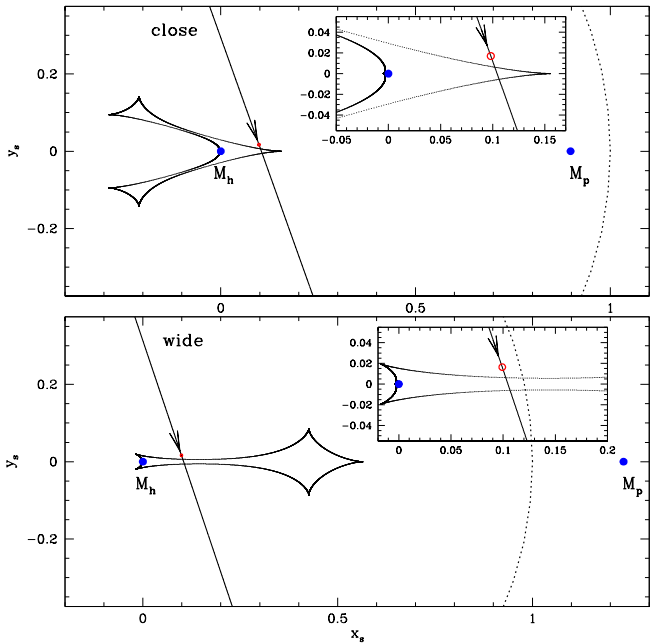}
\caption{Geometries of the close and wide binary lens systems.
The two lens components are marked as blue dots, while the red open circle represents the normalized source size.
The straight line with an arrow denotes the source trajectory, and the dotted black circle represents the Einstein ring.
The black closed curve denotes the caustic. \label{fig:jkasfig3}}
\end{figure}

We carry out a grid search for the binary lensing parameters $(s, q, \alpha)$ to find local $\chi^2$ minima using a downhill approach based on the Markov Chain Monte Carlo (MCMC) method.
The ranges of each parameter are $-1 \leqslant {\rm log} s \leqslant 1$, $-4 \leqslant {\rm  log} q \leqslant 0 $, and $0 \leqslant \alpha \leqslant 2\pi$ with (100, 100, 21) uniform grid steps, respectively.
During the grid search, $(s,q)$ are fixed and the other parameters are allowed to vary in the MCMC chain.
As a result, we find four local solutions of $(s,q)=(0.93,4.53\times10^{-3})$,  $(0.85,1.04\times10^{-2})$, $(1.18, 4.13\times10^{-3})$, and $(1.35, 1.38\times10^{-2})$, which indicate that the event has a well-known close/wide degeneracy $s \leftrightarrow 1/s$.
This close/wide degeneracy arises from the similarity in shape between the caustics induced by a close binary with $s<1$ and a wide binary with $s > 1$ (\citealt{griest1998}; \citealt{dominik1999}).
Figure \ref{fig:jkasfig1} shows the result of the grid search.
We then carry out an additional modeling in which the local solutions are set to the initial values and all parameters are allowed to vary.
From this, we find that each of the two close and wide solutions converges to $(s, q) = (0.90, 7.43\times 10^{-3})$ and $(1.23, 7.11\times 10^{-3})$, respectively.
The best-fit light curves of the close and wide models are shown in Figure \ref{fig:jkasfig2}.
The $\chi^2$ of the close model is smaller by $0.9$ than that of the wide model, and thus the event is severely degenerate.
The close and wide best-fit parameters with their 68\% uncertainty range from the MCMC method are listed in Table \ref{tab:jkastable1}, and the geometries of the two models are presented in Figure \ref{fig:jkasfig3}.

Even though the event timescale $(\te = 22\ \rm days)$ is not enough to measure the microlens parallax, we conduct the binary lens modeling with both the microlens parallax and lens orbital motion effects.
This is because that the orbital motion effect of the lens system can mimic the parallax signal (\citealt{batista2011}; \citealt{skowron2011}).
The microlens parallax is described by $\bm{\pie} = (\pien, \piee)$, in which the two components are given in equatorial coordinates \citep{gould1994}.
The lens orbital motion is described by two parameters $(ds/dt, d\alpha/dt)$, which represent the change rates of the binary separation and the orientation angle of the binary axis, respectively.
As we expected, we find that the parallax+orbital model is very weakly improved by $\delcs =2.6$ compared to the standard model, and the microlens parallax is not usefully constrained.

\begin{table}[t!]
\caption{Lensing parameters of the binary lens model.\label{tab:jkastable1}}
\centering
\begin{tabular}{lrr}
\toprule 
Parameter        &       Close                              &       Wide \\
\midrule
$\chi^2$/dof                   &   $1567.987/1611$              &       $1568.897/1611$  \\       
$t_0$ (HJD$^\prime$)      & $8563.8590 \pm 0.0216$     &       $8563.8840 \pm 0.0225$ \\
$u_0$                              &  $0.0982 \pm 0.0047$         &       $0.1020 \pm 0.0044$ \\
$\te$ (days)                     &  $22.2158 \pm 0.7249$        &       $21.8263 \pm 0.6839$ \\
$s$                                  &  $0.8980 \pm 0.0208$         &       $1.2342 \pm 0.0286$ \\
$q(10^{-3})$                    &  $7.4282\pm 1.5289$           &        $7.1095 \pm 1.5735$ \\
$\alpha$ (rad)                 &  $1.2303 \pm 0.0107$           &        $1.2486 \pm 0.0101$ \\
$\rho$                            &  $0.0034 \pm 0.0004$        &        $0.0031 \pm 0.0004$ \\
$f_{s,\rm ogle}$             &  $0.1498 \pm 0.0070$            &        $0.1550 \pm 0.0067$   \\
$f_{b,\rm ogle}$             &  $0.0236 \pm 0.0069$           &        $0.0184 \pm 0.0066$ \\
\bottomrule
\end{tabular}
\tabnote{ HJD$^\prime$ = HJD - 2450000.}
\end{table}

\begin{table}[t!]
\caption{Lensing parameters of the binary source model.\label{tab:jkastable2}}
\centering
\begin{tabular}{lr}
\toprule
Parameter                       &      1L2S   \\
\midrule
$\chi^2$/dof                         &   $2068.634/1611$            \\       
$t_{0,1}$ (HJD$^\prime$)   &  $8563.4507 \pm 0.0311$   \\
$u_{0,1}$                              &  $ 0.1213 \pm 0.0064$      \\
$t_{0,2}$ (HJD$^\prime$)  &  $8564.6231 \pm 0.0015$ \\
$u_{0,2}(10^{-3})$              &  $0.0221 \pm 0.2751$       \\
$\te$ (days)                          &  $22.8831 \pm 0.8945$        \\
$\rho_1$                                &  ...                                       \\
$\rho_2$                               &  $0.0060 \pm 0.0003$            \\
$q_{F}$                                &  $ 0.0636 \pm 0.0016$            \\
$f_{s,\rm ogle}$                  &  $0.1420 \pm 0.0079$  \\
$f_{b,\rm ogle}$                 &  $0.0314\pm 0.0078$  \\
\bottomrule
\end{tabular}
\end{table}

\subsection{Binary source model (1L2S)}
For a single lensing event induced by a binary source, the observed flux $F$ is the superposition of fluxes from the single lensing events of the two source stars (\citealt{griest1992}; \citealt{han1997}; \citealt{gaudi1998}),
\begin{equation}
F = \fsone A_1 + \fstwo A_2,
\end{equation}
where $\fsone$ and $\fstwo$ are the fluxes of the primary $(\rm S_1)$ and companion $(\rm S_2)$ sources, respectively, and $A_1$ and $A_2$ are the lensing magnifications by the primary and companion sources.
Thus, the total magnification for the binary source lensing \citep{hwang2013} is represented by
\begin{equation}
A = {A_{1} \fsone + A_2 \fstwo\over{\fsone + \fstwo}} = {A_1 + q_{F} A_2\over{1 + q_{F}}},
\end{equation}
where $q_{F}=\qflux$, and
\begin{equation}
A_{i} = {u^{2}_{i} + 2\over{u_{i} \sqrt{u^{2}_{i} + 4}}}; \quad u_i = \left[u^{2}_{i} + {\left({t-t_{0,i}\over{\te}}\right)^2}\right].
\end{equation}
In order to mimic the anomaly induced by the binary lens,  as shown in Figure \ref{fig:jkasfig2}, one of the two sources has to pass close to the lens, which means that its $\uo$ is very small, thus making it highly magnified.
For the 1L2S modeling, we need 8 parameters: the single lens parameters for the two sources $\rm S_1$ and $\rm S_2$, $(t_{0,1}, u_{0,1}, \rho_1)$ and $(t_{0,2}, u_{0,2}, \rho_2)$, $\te$, and the flux ratio of the two sources $q_{F}$ (\citealt{griest1992}; \citealt{jung2017}).
The $(\tzero,\uo, \te, \rho)$ of the binary lens solution are set to the initial values of the parameters $(t_{0,1}, u_{0,1}, \te, \rho_2)$, while we set the initial values of the parameters $(t_{0,2}, u_{0,2}, q_{F})$ by considering the peak time and the magnification of the short duration anomaly that was obtained from the binary lens.
The best-fit parameters for the binary source model are presented in Table \ref{tab:jkastable2}.
From the result of the 1L2S modeling, we find that the $\delcs$ between 2L1S and 1L2S models is $\delcs = 500.6$.
This means that the event OGLE-2019-BLG-0362/KMT-2019-BLG-0075 is caused by a binary lens system.

\section{Angular Einstein radius}
In order to measure the angular Einstein radius $\thetae$, one should measure the angular source radius and so obtain $\thetae = \thetas/\rho$.
As mentioned in Section~2, KMT data were taken in  the $I-$ and $V-$bands to measure the source color.
The measured instrumental color and magnitude of the source are $(V-I)=3.35$ and $I=20.06$, which are obtained from a regression and the source flux of the best-fit model, respectively.
The angular source radius is estimated from the intrinsic color and magnitude of the source, in which they are obtained from the offset between the red giant clump and the source positions on the instrumental CMD,
\begin{equation}
\Delta (V-I, I) = (V-I, I)_0 - (V-I, I)_{\rm cl,0}.
\end{equation}
We thus construct the KMTC CMD from the pyDIA pipeline.
From the CMD, we find that the color and magnitude of the clump are $(V-I, I)_{\rm cl}=(3.61, 17.44)$.
However, the measured instrumental source color was obtained from three very low-magnified $V$ band points on the wing of the light curve, and the extinction toward the event is high as $A_I=3.04$, making it doubtful that the color is reasonable.
In addition, unfortunately, there was no magnified $V$ band data for OGLE.
Hence, we combine the KMTC CMD and CMD constructed from the Galactic bulge images taken from the \textit{Hubble} Space Telescope (\textit{HST}) \citep{holtzman1998}.
The combination of the two CMDs is performed by calibrating the positions of the clumps on each CMD.
Figure 4 shows the combined CMD.
From the CMD, we find that the source color is $(V-I)=3.27 \pm 0.13$, which is estimated by taking the average of the calibrated $HST$ stars that are in the ranges of $1.1 \lesssim (V-I)_0 \lesssim 1.5$ and $17.5 \lesssim I_0 \lesssim 17.6$.
The offset is thus $\Delta(V-I, I)=(-0.34, 2.62)$.
We adopt the intrinsic color and magnitude of the clump: $(V-I)_{\rm cl,0}=1.06$ from \citet{bensby2011} and $I_{\rm cl,0}=14.37$ from \citet{nataf2013}.
As a result, we find that the intrinsic color and magnitude of the source are $(V-I, I)_0 = (0.72 \pm 0.13, 16.99 \pm 0.01)$.
This indicates that the source is a G-type turn-off star or a G-type subgiant.
The intrinsic $(V-I)_0$ source color is converted to $(V-K)_0$ color using the color-color relation of \citet{bessell1988}, and then adopting the $(V-K)_0$ to the color-surface brightness relation of \citet{kervella2004}, we obtain the angular source radius $\thetas = 1.40 \pm 0.19\, \rm \mu as$ for the close and wide models.
We then estimate the angular Einstein radii for the close and wide models as
\begin{equation}
\thetae = \thetas/\rho = \left\lbrace 
\begin{array}{ll}
0.416 \pm 0.082\, \textrm{mas} & \textrm{(close)} \\
0.443 \pm 0.065\, \textrm{mas} & \textrm{(wide)}.
\end{array} \right.
\end{equation}
The relative lens-source proper motion is estimated as
\begin{equation}
\murel = \thetae/\te = \left\lbrace 
\begin{array}{ll}
6.85 \pm 1.32\, \textrm{mas\, yr$^{-1}$} & \textrm{(close)} \\
7.41 \pm 1.08\, \textrm{mas\, yr$^{-1}$} & \textrm{(wide)}.
\end{array} \right.
\end{equation}

\begin{figure}
\centering
\includegraphics[width=90mm]{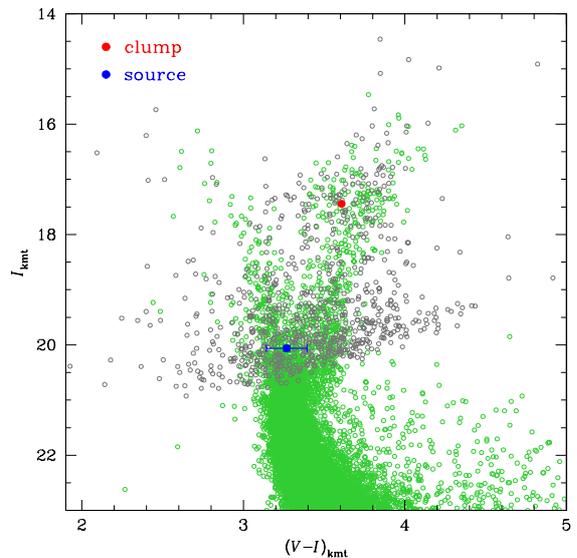}
\caption{Combined CMD by KMTC and \textit{HST} CMDs.
The KMTC CMD is constructed by the pyDIA reductions and is marked as gray dots, while the \textit{HST} CMD is constructed from the Galactic bulge images taken by the \textit{HST} and is marked as green dots.
The red and blue dots represent the positions of the red giant clump and source star, respectively.\label{fig:jkasfig4}}
\end{figure}

\section{Physical lens properties}
For the event OGLE-2019-BLG-0362, the microlens parallax was not measured, and thus we cannot directly measure the physical parameters of the planetary lens system.
In this case, we perform a Bayesian analysis with the measured $\thetae$ and $\te$ in order to constrain the physical lens parameters.
The Bayesian analysis assumes that all stars have an equal probability to host a planet of the measured mass ratio (\citealt{bhattacharya2021}; \citealt{vandorou2020}).
For the Bayesian analysis, we follow the procedures of \citet{jung2018}, but we use the new Galactic model constructed by \citet{jung2021}.
The new Galactic model of \citet{jung2021} includes the bulge mean velocity taken from stars in the Gaia catalog, disk density profile and disk velocity dispersion from the Robin-based model \citep{bennett2014}, while the remaining parameters, including the bulge mean velocity, the bulge density profile, and mass function, are the same as those of \citet{jung2018}.
Figure 5 shows the posterior probability distribution of the mass and distance of the host star derived from the Bayesian analysis for the two models.
Due to the close/wide degeneracy, each model has a different $\thetae$.
Thus, we find that the estimated masses of the host and planet are
\begin{equation}
\label{eqn:mass}
(M_{\rm host}, M_{\rm p})= \left\lbrace
\begin{array}{ll}
(0.42^{+0.34}_{-0.23}\, \rm M_\odot,\, 3.26^{+0.83}_{-0.58}\, M_{\rm J})\, \textrm{(close)} \\
(0.45^{+0.33}_{-0.24}\, \rm M_\odot,\,  3.34^{+0.78}_{-0.58}\, M_{\rm J})\, \textrm{(wide)},
\end{array} \right.
\end{equation}
where $M_{\rm p}=qM_{\rm host}$, and the distance to the lens is
\begin{equation}
\label{eqn:distance}
\dl= \left\lbrace
\begin{array}{ll}
5.83^{+1.04}_{-1.55}\, \rm kpc\, \textrm{(close)} \\
5.72^{+1.03}_{-1.57}\, \rm kpc\, \textrm{(wide)}.
\end{array} \right.
\end{equation}
The projected star-planet separations of the close and wide models are
\begin{equation}
\label{eqn:distance}
a_\perp = s\dl\thetae = \left\lbrace
\begin{array}{ll}
2.18^{+0.58}_{-0.72}\,\rm AU\, \textrm{(close)} \\
3.13^{+0.73}_{-0.98}\,\rm AU\, \textrm{(wide)},
\end{array} \right.
\end{equation} 
respectively.
\begin{figure}
\centering
\includegraphics[width=90mm]{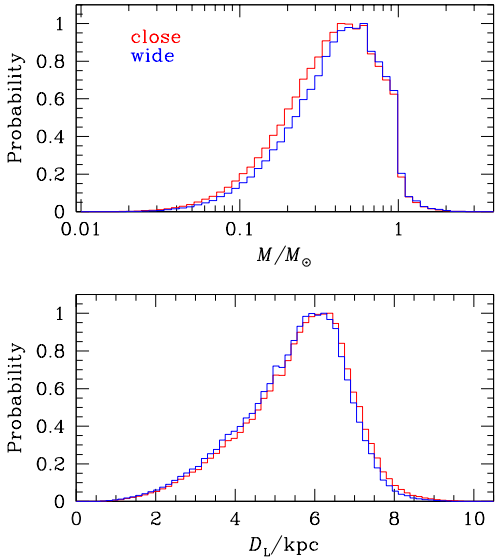}
\caption{Posterior probability distributions of the mass and distance of the host star estimated from the Bayesian analysis based on the close and wide models.
The red and blue curves represent the distributions for the close and wide models, respectively.\label{fig:jkasfig5}}
\end{figure}
The physical values of the lens system are the median values of the Bayesian posterior distributions, and their uncertainties indicate the 68\% confidence intervals (i.e., $1\sigma$ errors) of the distributions.
Considering the snow line of $a_{\rm snow}=2.7(M/M_\odot)$\citep{kennedy2008}, the planet is orbiting beyond the snow line of the M dwarf star.
However, it could also be a K dwarf star because the host star has the mass  of $0.19 \sim 0.78\, M_\odot$ at $1\sigma$ level.
Due to the severe close/wide degeneracy, the physical lens parameters for the two models are almost the same.
Considering of the estimated lens distance by the Bayesian analyses and $\murel \simeq 7\, \rm mas\, yr^{-1}$, the lens system is likely to be located at the disk.
The physical lens parameters for the two models are presented in Table \ref{tab:jkastable3}.

\begin{table}[t!]
\caption{Physical lens parameters.\label{tab:jkastable3}}
\centering
\begin{tabular}{lrr}
\toprule
Parameter        &       Close                              &       Wide \\
\midrule
$M_{\rm host}$ $(M_\odot)$             & $0.42^{+0.34}_{-0.23}$      &   $0.45^{+0.33}_{-0.24}$ \\
$M_{\rm p}\, (M_{\rm J})$                   &  $3.26^{+0.83}_{-0.58}$     &    $3.34^{+0.78}_{-0.58}$\\
$D_{\rm L}$ (kpc)                                &   $5.83^{+1.04}_{-1.55}$   &    $5.72^{+1.03}_{-1.57}$ \\
$a_{\perp}$ (au)                                  &   $2.18^{+0.58}_{-0.72}$   &   $3.13^{+0.73}_{-0.98}$\\
$\mu_{\rm rel}\, (\rm mas\ yr^{-1})$ &   $6.86 \pm 1.32$              &   $7.42 \pm 1.08$ \\
\bottomrule
\end{tabular}
\tabnote{The physical parameters are obtained by the Bayesian analyses.
The representative values are chosen as the median values of the Bayesian posterior distributions, and their uncertainties indicate the 68\% confidence intervals of the distributions.}
\end{table}

\section{Summary}
We reported a planetary system discovered from the analysis of the microlensing event OGLE-2019-BLG-0362/KMT-2019-BLG-0075.
The event has a distinctive anomaly feature near the peak of the light curve, thus it looks like a typical 2L1S event.
However, the anomaly was covered by only three KMT data points due to its short duration of $0.4\, \rm days$, thus making it difficult to securely insist that this is a 2L1S event. 
We thus conducted two kinds of modelings with 1L2S and 2L1S, which can produce the short duration anomaly.
As a result, it is found that the event is induced by the 2L1S system because the $\chi^2$ of the 1L2S model is much larger than that of the 2L1S model by $\delcs=501$.
The binary lensing solution is subject to the close/wide degeneracy, and this degeneracy is very severe because of $\delcs < 1$ between the close and wide models.
Due to a relatively short event timescale of $\te = 22\, \rm days$, the microlens parallax was not measured.
We thus carried out a Bayesian analysis, and from this, it is found that the lens is composed of $(M_{\rm host}, M_{\rm p}) = (0.42^{+0.34}_{-0.23}\, \rm M_\odot,\, 3.26^{+0.83}_{-0.58}\, M_{\rm J})$ for the close model, while for the wide model it is composed of $(M_{\rm host}, M_{\rm p})=(0.45^{+0.33}_{-0.24}\, \rm M_\odot,\,  3.34^{+0.78}_{-0.58}\, M_{\rm J})$.
The distances to the lens for the close and wide models are $\dl = 5.83^{+1.04}_{-1.55}\, \rm kpc$ and $5.72^{+1.03}_{-1.57}\, \rm kpc$, respectively.
The Bayesian distributions of the lens distance and relative lens-source proper motion of $\murel \simeq 7\, \rm mas\, yr^{-1}$ indicate that the lens is likely to be located at the disk.
Due to $a_\perp > 2\, \rm AU$, it is found that the planet orbits beyond the snow line of an M dwarf or a K dwarf.
The relative lens-source proper motion is $\murel \simeq 7\, \rm mas\, yr^{-1}$, and thus the lens will be separated from the source by $\simeq 70\, \rm mas$ in 2029, at which point one can measure the lens flux from adaptive optics of next-generation 30 m telescopes.

\acknowledgments
Work by S.-J. Chung was was supported by the Korea Astronomy and Space Science Institute under the R\&D program (Project No. 2022-1-830-04) supervised by the Ministry of Science and ICT.
J.C.Y. acknowledges support from N.S.F Grant No. AST-2108414.
Work by C.H. was supported by the grants of National Research Foundation of Korea (2020R1A4A2002885 and 2019R1A2C2085965).
This research has made use of the KMTNet system operated by the KASI and the data were obtained at three sites of CTIO in Chile, SAAO in South Africa, and SSO in Australia.

\end{document}